\documentclass[conference]{IEEEtran}
\IEEEoverridecommandlockouts
\usepackage{cite}
\usepackage{amsmath,amssymb,amsfonts}
\usepackage{algorithmic}
\usepackage{graphicx}
\usepackage{textcomp}
\usepackage{xcolor}
\usepackage{booktabs}
\def\BibTeX{{\rm B\kern-.05em{\sc i\kern-.025em b}\kern-.08em
    T\kern-.1667em\lower.7ex\hbox{E}\kern-.125emX}}
\begin{document}

\title{CRPE: Expanding The Reasoning Capability of Large Language Model for Code Generation
}

\author{\IEEEauthorblockN{Ningxin Gui\IEEEauthorrefmark{1}, 
Qianghuai Jia\IEEEauthorrefmark{2}, 
Feijun Jiang\IEEEauthorrefmark{2}, 
Yuling Jiao\IEEEauthorrefmark{1}, 
dechun wang\IEEEauthorrefmark{2}, 
Jerry Zhijian Yang\IEEEauthorrefmark{1}
}
\IEEEauthorblockA{\IEEEauthorrefmark{1} School of Mathematics and Statistics \\
Wuhan University \\
Email: gui-ningxin@whu.edu.cn
}
\IEEEauthorblockA{\IEEEauthorrefmark{2}Alibaba International Digital Commerce}
}



\maketitle

\begin{abstract}
We introduce CRPE (Code Reasoning Process Enhancer), an innovative three-stage framework for data synthesis and model training that advances the development of sophisticated code reasoning capabilities in large language models (LLMs). Building upon existing system-1 models, CRPE addresses the fundamental challenge of enhancing LLMs' analytical and logical processing in code generation tasks. Our framework presents a methodologically rigorous yet implementable approach to cultivating advanced code reasoning abilities in language models.
Through the implementation of CRPE, we successfully develop an enhanced COT-Coder that demonstrates marked improvements in code generation tasks. 
Evaluation results on LiveCodeBench (20240701-20240901) demonstrate that our COT-Coder-7B-StepDPO, derived from Qwen2.5-Coder-7B-Base, with a pass@1 accuracy of 21.88, exceeds all models with similar or even larger sizes. Furthermore, our COT-Coder-32B-StepDPO, based on Qwen2.5-Coder-32B-Base, exhibits superior performance with a pass@1 accuracy of 35.08, outperforming GPT4O on the benchmark.
Overall, CRPE represents a comprehensive, open-source method that encompasses the complete pipeline from instruction data acquisition through expert code reasoning data synthesis, culminating in an autonomous reasoning enhancement mechanism.
\end{abstract}

\begin{IEEEkeywords}
Code generation, reasoning enhancing
\end{IEEEkeywords}

\section{Introduction}
\label{Introduction}
\begin{figure*}[htbp]
\begin{center}
\centerline{\includegraphics[width=\textwidth]{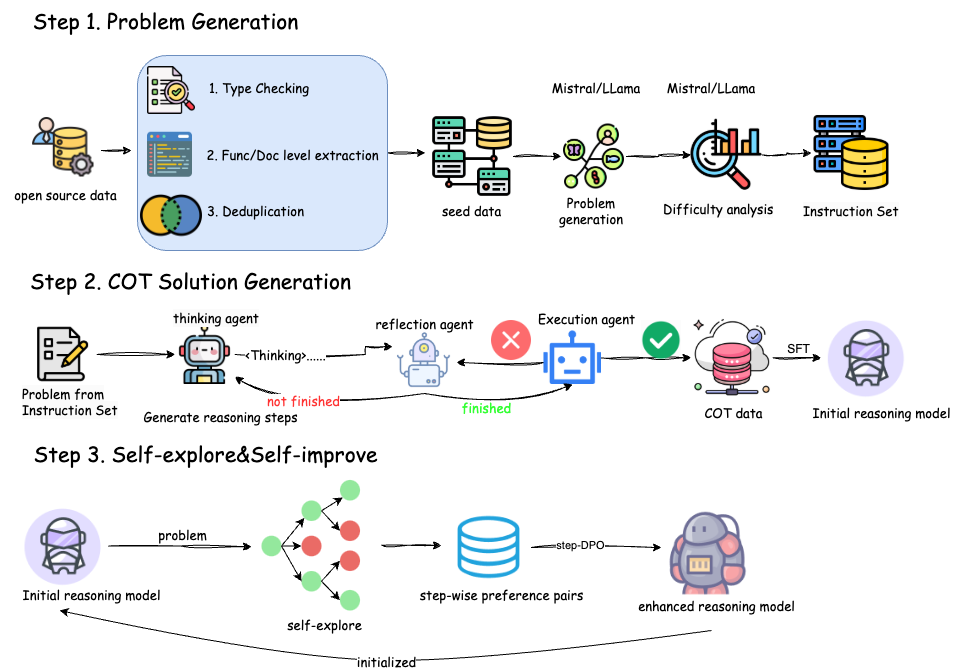}}
\caption{The overview of CRPE, which comprises three steps: 
 (i) synthesizing large-scale code problems and (ii) synthesizing high-quality code reasoning datas for supervised finetuning and (iii) prompting code reasoning model derived from the second step to sample reasoning paths and achieve self-improve.} 
\label{fig:my_label1} 
\end{center}
\end{figure*}
Code generation, also referred to as program synthesis, aims to automatically generate source code that adheres to established programming specifications, which are typically articulated in natural language.
In recent years, the coding capabilities of large language models have advanced significantly \cite{hui_qwen25-coder_2024,deepseek-ai_deepseek-coder-v2_2024}, allowing them to help users address practical coding challenges effectively.

Nevertheless, these models often produce suboptimal results when confronted with complex coding problems. This is primarily because such problems necessitate intricate reasoning processes to derive solutions, which require the models to possess System 2 thinking capabilities. Prior to the introduction of OpenAI O1, large-language models predominantly exhibited System 1 capabilities in the context of code generation tasks. Although methodologies such as Chain of Thought (COT)\cite{Wei-CoT-2201-11903} and Tree of Thought (TOT)\cite{yao2023treethoughtsdeliberateproblem} exist to facilitate step-by-step reasoning by the models, these approaches are contingent upon and constrained by the models' inherent reasoning abilities. 
A recent series of reasoning models has demonstrated promising performance on challenging mathematical and coding tasks. However, the optimal approach for training such reasoning models remains unclear. Consequently, investigating methods to enhance the reasoning capabilities of these models is of paramount importance in advancing the field of solving complex code generation.

Current research has made significant strides in improving the mathematical reasoning capabilities of large-language models\cite{chen2024alphamathzeroprocesssupervision, zhang2024llamaberrypairwiseoptimizationo1like}.
Many efforts aimed at improving coding capabilities focus on generating challenging and high-quality code problems and solutions for supervised fine-tuning, overlooking the reasoning process that leads to those code answers\cite{luo2023wizardcoderempoweringcodelarge, yu2024wavecoderwidespreadversatileenhancement, wei2024magicoderempoweringcodegeneration}. There are also some methods based on reinforcement learning \cite{dou2024stepcoderimprovecodegeneration, shojaee2023executionbasedcodegenerationusing, liu2023rltfreinforcementlearningunit} that allow models to improve their coding abilities through self-exploration, these approaches primarily emphasize generating the correct final answer rather than focusing on the code reasoning process. However, the reasoning process underlying code generation is equally crucial for enhancing a model's coding abilities.
To address this challenge, we propose a novel method called Code Reasoning Process Enhancement (CRPE), which can effectively improve the code reasoning capabilities of LLMs.
This approach aims to bridge the gap with O1 and bring improvements in code reasoning, focusing on the step-by-step thought process that leads to high-quality code solutions.

In our approach, considering the scarcity of high-quality and high-difficulty data, we first establish a code instruction generation pipeline, which includes the collection and filtering of open-source data as well as instruction synthesis based on LLM. Ultimately, this pipeline enables us to create a substantial number of high-quality and challenging code problems. Next, considering the lack of high-quality code reasoning data, we design a multi-agent framework that leverages powerful system-1 models to synthesize high-quality code reasoning data. This data is then used for SFT, allowing LLMs to learn the code reasoning thoughts of experts. Finally, considering the high cost associated with synthesizing expert code reasoning data, we develop a tree search-based self-exploration and enhancement method. This enables the code reasoning model to explore both correct and incorrect reasoning steps, facilitating self-improvement through Step-DPO. 

In summary, CRPE employs a three-stage data synthesis and training method:
\begin{enumerate}
  \item We design a simple yet effective data synthesis method to generate competition-level programming problems.
  \item We develop a sophisticated multi-agent framework to synthesize high-quality Chain of Thought (CoT) data. This data serves as a cold start instruction set, guiding the model's fundamental cognitive processes and output structures.
  \item We leverage the reasoning code model trained using the data generated in the second stage. We employ our optimized tree search algorithm to sample new reasoning process data. A noteworthy byproduct of this tree search process is the generation of abundant step-wise preference pairs. These pairs are subsequently leveraged to further refine the model through step-wise Direct Preference Optimization (step-DPO). This stage can be executed iteratively, allowing for continuous refinement and expansion of the model's reasoning capabilities.
\end{enumerate}

Experimental results on LiveCodeBench demonstrate that our approach effectively enhances the code generation abilities of LLM by improving its code reasoning capabilities.

Our contributions are as follows: \begin{itemize}
    \item We propose the CRPE method, a three-stage data synthesis and training method that effectively enhances the code reasoning capabilities of models. 
    \item We propose a synthesis pipeline for code instruction data and construct a large-scale, high-quality code problems dataset.
    \item We design a multi-agent framework and construct a high-quality code-COT SFT dataset.
    \item We design a novel synthesis method for collecting step-wise code-COT preference data for step-DPO.
    \item Experiments show that CRPE can effectively enhance the large-language model's code generation capabilities.
\end{itemize}
\section{Related Work}
\textbf{CodeLLM}. LLMs have demonstrated impressive capabilities in code generation tasks after being trained on extensive code datasets.\cite{hui2024qwen25codertechnicalreport, guo2024deepseekcoderlargelanguagemodel, rozière2024codellamaopenfoundation} Since high-quality data is hard to get, \textbf{synthetic data} becomes a crucial approach to enhance the code capabilities of models. The key point of this methodology lies in leveraging powerful foundation models to generate high-quality instructional data and corresponding code solutions. This process involves defining certain criteria for instruction filtering and utilizing compilers to ensure the correctness of the generated code. Synthetic data can be used to improve the code capabilities of other weaker models or to achieve self-improve. \cite{grattafiori2024llama3herdmodels, lozhkov2024starcoder2stackv2}Llama3 and StarCoder2 synthesize difficult and high-quality code data by building a complete data synthesis pipeline that starts from code databases. \cite{luo2023wizardcoderempoweringcodelarge, wei2024magicoderempoweringcodegeneration}Evol-Instruct and OSS-instruct design special prompt to improve instruction data. Finally the synthesized code data is used in the form of QA pairs for \textbf{SFT}. 
Some other works aim to enhance the code capabilities of LLM through \textbf{reinforcement learning}. Since code tasks can obtain execution information through a compiler, it is possible to collect or generate test cases to execute and the execution result can be used as rewards to train the policy model\cite{wang2022compilableneuralcodegeneration, le2022coderlmasteringcodegeneration, shojaee2023executionbasedcodegenerationusing, liu2023rltfreinforcementlearningunit}. \cite{dou2024stepcoderimprovecodegeneration}StepCoder breaks down the code generation process and completes the answer incrementally, reducing the difficulty for the model to explore the correct solution. \cite{zhang2024codedpoaligningcodemodels} CodeDPO optimizes the correctness and efficiency of solutions through preference learning. But these works only teach LLM the correct answers, rather than how to reason to arrive at those answers.

\textbf{Enhancing Code Reasoning capabilities of LLM}. Recently, the reasoning capabilities of LLM have been further enhanced. Notable works include OpenAI O1\cite{openaio1}, QwQ\cite{qwenqwq}, DeepSeek-R1-Lite. These reasoning LLMs have demonstrated outstanding performance in complex mathematical reasoning and code generation tasks, but it is unclear how to achieve such a coding reasoning model. Some works \cite{dai2024processsupervisionguidedpolicyoptimization, zhang2024o1codero1replicationcoding} implement code reasoning models through training process reward model(PRM) and reinforcement learning (RL). These methods need to train more than one model, increasing the complexity of the training process. Our work offers a new method to enhance code reasoning capabilities of LLMs with the help of strong system-1 LLMs.

\section{methodology}
In this section, we present our CRPE in detail to enhance code reasoning capabilities of LLMs. We first introduce some background on post-training in subsection 3.1. Then, in subsection 3.2, we explain the methods used to collect and generate the code problems that we will use for training later on. In subsection 3.3, we present the multi-agent framework we designed to generate Code-COT SFT data, which enables the model to adopt paradigms for code reasoning and strengthen its code reasoning capabilities. Finally, in subsection 3.4, we illustrate a pipeline for generating reasoning data to achieve self-improve.

\subsection{Preliminaries}

The post-training of large language models typically involves two steps: supervised fine-tuning (SFT) and reinforcement learning based on human feedback (RLHF). SFT is usually conducted on Question-Answer(QA) pairs. 

RLHF consists of two stages: training the reward model and training the policy model. The entire process is quite complex.
\cite{rafailov2024directpreferenceoptimizationlanguage} proposed DPO, which simplifies the training process of RLHF by directly training the policy model using preference data.

To further enhance the model's reasoning capabilities, it is essential to pay more attention to the details of each reasoning step during training. For hard-to-think question $x$, the answer $y$ might also include all the reasoning steps. So $y$ can be decomposed into a sequence of reasoning steps $s_1,...,s_n$, $s_i$ is the i-th reasoning step. To achieve more fine-grained preference learning, \cite{lai2024stepdpostepwisepreferenceoptimization} proposed Step-DPO, which uses step-wise preference data pairs for training. Specifically, the goal of Step-DPO is to maximize the probability of the next correct reasoning step while minimizing the probability of the next incorrect reasoning step, given the problem and the partial correct reasoning steps.

\subsection{Code Problem Data Preparation}

\begin{figure}[htbp]
    \centering
    \includegraphics[width=0.4\textwidth]{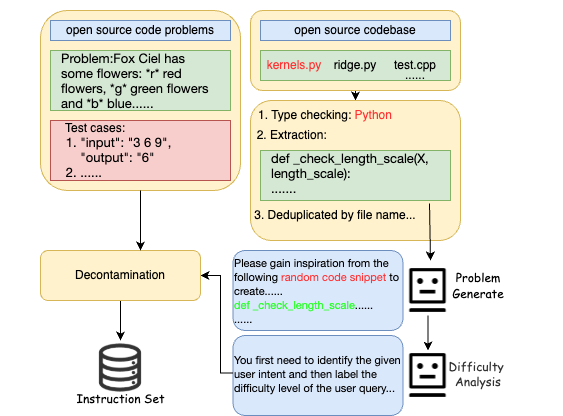}
    \caption{Code Problem Pipeline.} 
    \label{fig:my_label_data_prepare} 
\end{figure}

Considering that the majority of open-source code data may have already been used during the pre-training of LLMs, the associated problems may not require complex reasoning for the models to arrive at correct answers. As such, further training on these problems does not effectively enhance the models' code reasoning capabilities. Therefore, in addition to collecting open-source code instructions, we also construct more challenging code instructions through data synthesis.

\subsubsection{open-source data collection}
We collect some open-source code problems. This data primarily consists of difficult programming problems from Codeforces or LeetCode, along with the corresponding test cases and potential correct answers. The test cases are used to verify the correctness of the generated code by LLM. When the model is unable to generate correct code, the possible correct answers can assist the model in analyzing coding issues and making corrections. Due to the difficulty in obtaining this high-quality data, we ultimately gather several thousand code instructions along with their respective test cases and potential correct code answers.

\subsubsection{code problem synthesis}
To enhance the diversity and complexity of the code instructions, we also refer to the methods of OSS-Instruct\cite{wei2024magicoderempoweringcodegeneration} and Evol-Instruct\cite{luo2023wizardcoderempoweringcodelarge}. We extract code snippets from the seed data of stackv2\cite{lozhkov2024starcoder2stackv2} and perform deduplication by file name and function name. The extracted code snippets are used to prompt LLMs (such as Llama or Mistral Large) to generate new code instructions, and we utilize Evol-Instruct to optimize these code instructions. Finally, we filter the synthesized instructions through LLMs. The main filtering rules included: (1) whether the instruction has a clear intent, meaning it requires writing a piece of code to achieve a specific goal; (2) whether the instruction is challenging, prompting LLMs to judge whether the problem is difficult and requires reasoning to arrive at the answer; and (3) whether the problem is self-contained, meaning it can be solved using only Python's standard library to avoid execution errors due to environmental dependencies. Ultimately, we generate 2 million code instruction data, of which only a portion is used in latter training.

\subsubsection{decontamination}
To ensure that our model does not produce inflated results due to the leakage of the test set, we perform a decontamination on all the code problems obtained. The specific method is similar to that used in Qwen2.5-Coder\cite{hui2024qwen25codertechnicalreport}, whereby any data in the training set that has a 10-gram overlap with the test data is removed.

\subsection{Code-COT Maker Based On Multi-Agent Framework}
Code-COT Maker consists of three agents (Thinking Agent, Reflection Agent, and Execution Agent) that operate in a workflow. The Thinking Agent receives coding problems and engages in step-by-step reasoning to arrive at a final answer. The Reflection Agent's task is to analyze whether the reasoning steps of the Thinking Agent are correct and to decide whether the next step should be for the Thinking Agent to continue reasoning or to output the code answer. If the Reflection Agent decides to output the code answer, the Execution Agent will use a compiler and LLM-as-Critic to determine whether the code answer is correct. If the code answer is correct, the process concludes; if the code answer is incorrect, the Execution Agent will pass the execution results to the Reflection Agent, which will analyze the coding errors in conjunction with the current answer and prompt the Thinking Agent to generate a new answer. A threshold for the maximum number of execution feedback attempts will be pre-set to prevent the entire workflow from becoming non-terminating. Ultimately, the reasoning paths deemed correct by the Execution Agent will be retained for training.

\subsubsection{Thinking Agent}

\begin{figure}[ht]
    \centering
    \includegraphics[width=0.4\textwidth]{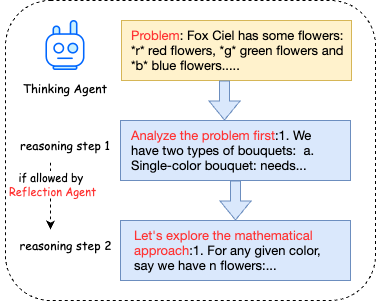}
    \caption{The Thinking Agent Illustration.} 
    \label{fig:my_label11} 
\end{figure}

The thinking agent is powered by LLMs. Its task is to engage in step-by-step reasoning to generate the final code answer and to rectify incorrect code answers. We designed special system prompts that encourage the Thinking Agent to undertake multi-step reasoning before providing an answer, without limiting the number of reasoning steps or the specific content of each step; these are determined by the coding problem and the model itself. If the final answer provided by the Thinking Agent is incorrect, it will receive a code error analysis report from the Reflection Agent and will optimize the code based on that report until it passes the checks from the Execution Agent or reaches the maximum number of allowed checks.

\subsubsection{Reflection Agent}

\begin{figure}[ht]
    \centering
    \includegraphics[width=0.4\textwidth]{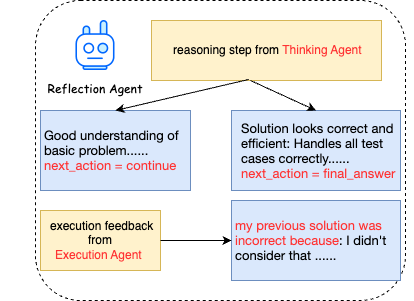}
    \caption{The Reflection Agent Illustration.} 
    \label{fig:my_label2} 
\end{figure}

The Reflection Agent is also powered by LLMs. It needs to analyze the reasoning content of each step provided by the Thinking Agent and determine whether it is correct, thereby deciding whether to allow the Thinking Agent to continue reasoning or to hand the output over to the Execution Agent for evaluation. When the Execution Agent determines that the generated answer is incorrect, the Reflection Agent will analyze the reasons for the coding errors by combining the execution results from the Execution Agent with the code answers by the Thinking Agent. This analysis is then summarized into a code error analysis report, which is sent back to the Thinking Agent for it to regenerate the answer. If, after several checks by the Execution Agent, the code still contains errors and our data includes correct code answers, the correct answers will be provided to the Reflection Agent to offer more accurate reflections. However, the Reflection Agent does not directly provide the correct answer; rather, it analyzes the reasons for the current answer's errors based on the correct answers.

\subsubsection{Execution Agent}
\begin{figure}[ht]
    \centering
    \includegraphics[width=0.4\textwidth]{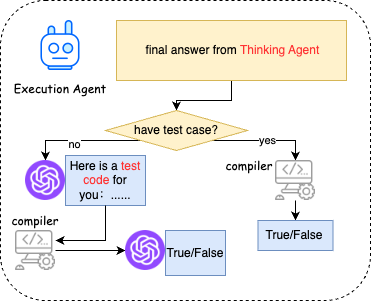}
    \caption{The Execution Agent Illustration.} 
    \label{fig:my_label3} 
\end{figure}
The task of the Execution Agent is to verify whether the code generated by the Thinking Agent is correct. It consists of three components: the Compiler, the Test Generator, and the Result Checker. If the current coding problem includes executable test cases, the Execution Agent will directly execute the code along with the test cases and provide a signal indicating whether the outcome is correct or incorrect. If the execution result is incorrect, it will relay specific error information to the Reflection Agent. If the coding problem does not come with test cases, the Test Generator will create test cases based on the coding problem and the code answer, organizing them into an executable format. The generated test cases are then executed using the Compiler, and the execution results are passed to the Result Checker to determine the correctness of the code. Both the Test Generator and the Result Checker are driven by LLMs. 

\subsection{Reasoning Enhancer by self-improve}
\begin{figure}[ht]
    \centering
    \includegraphics[width=0.4\textwidth]{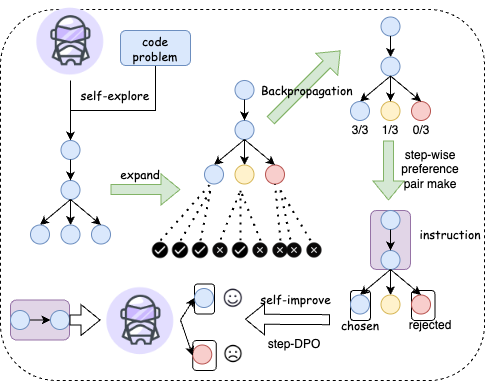}
    \caption{self-explore\&self-improve.} 
    \label{fig:my_label4} 
\end{figure}
Although training the model using Code reasoning data synthesized with powerful system-1 model can enhance the model's code reasoning capabilities, generating such synthetic data is very costly and will ultimately be limited by the capabilities of the system-1 model. Therefore, we also explore the self-improvement of the Code-COT model. We propose an improved tree-search algorithm to collect stepwise preference data. Using the collected stepwise preference data, we optimize the code reasoning ability of the model through step-DPO. 

Next, we introduce our improved tree search algorithm for synthesizing single-step preference data. However, checking the incorrect step within a reasoning path is quite challenging, as we can only determine whether a path is erroneous based on the execution results at the final answer node, making it difficult to pinpoint which specific step is at fault. Based on this analysis, we designed an improved tree search to assist in generating single-step preference data.

Our tree search algorithm consists of four iterative steps: select, expand, simulate, and backpropagation. Similarly to the approach used in \cite{xin2024deepseekproverv15harnessingproofassistant}, we combine the expansion and simulation steps. After selecting a node, a complete path is generated along that node, and the correctness of the generated path is assessed through code execution.

If a node leads to a correct answer node, it is classified as an "accepted" node. Conversely, if a node has failed to yield a correct answer after sampling more than  `max\_path\_num` paths, it is classified as a "rejected" node, indicating that it is erroneous or that generating a correct answer from that node is difficult. We use the hyperparameter `max\_path\_num` to limit the width of the code tree and `max\_depth\_num` to restrict the depth of the code tree.

The specific details of the tree search are as follows:

\textbf{Selection}: We perform a level-order traversal starting from the root node. If the nodes at the current level have been fully explored, meaning that the number of paths originating from these nodes has reached `max\_path\_num`, we will select node from the "accepted" child nodes based on which the generated paths are not all correct, until reaching an answer node or the exploration depth reaches `max\_depth\_num`.

\textbf{Expansion}: From the selected node, we generate the remaining paths and perform multiple samplings until the number of sampled paths reaches `max\_path\_num`. The answer code in each path is executed for verification. Then we mark any node whose final answers are all incorrect as a "rejected" node; otherwise, it is classified as an "accepted" node.

\textbf{Backpropagation}: Retain all paths generated from the selected node and update the counts of paths and correct paths for all nodes originating from the selected node.

For each "accepted" node in the final generated code reasoning tree, we select "accepted" child nodes and "rejected" child nodes, or "accepted" child nodes with significantly different accuracy rates, to construct preference data pairs. For each node, we select two child nodes to form one preference pair for further step-DPO.

\section{experiment}
In this section, we first introduce the experimental setup in Sec. 4.1. Then, we present the main results in Sec. 4.2. In Sec.4.3, we analyze the effectiveness of code reasoning data and evaluate the performance of our COT-Coder on other benchmarks. 
\subsection{Experimental Setup}
\textbf{Code Problem Preparation} During the data synthesis phase, we use Llama-3-70B-Instruct for problem generation and Mistral-large for difficulty analysis. In the data decontamination process, we compare our code problems with all the code problems in LiveCodeBench one by one and remove similar data. Using our Code Problem Pipeline, we finally collect 4,749 open-source code problems and extract 6k code problems from the synthesized code problem dataset. These code problems are utilized for code reasoning process generation and later training. 

\textbf{Code-COT data generation} We use the code problem data obtained above and employ Claude-3.5-sonnet to act as the three agents within the Code-COT Maker, utilizing the powerful System-1 model to synthesize high-quality code reasoning data. During the analysis of the outputs from the Thinking Agent and the Reflection Agent, we find that the Reflection Agent is unable to provide effective reflections on the Thinking Agent's one small reasoning step, often responding with "the reasoning step now is correct, please continue to reason." Considering that these intermediate reflections don't aid in reasoning and waste tokens, we eliminate these intermediate reflection processes, retaining only the reflections from the Reflection Agent after receiving feedback from the Execution Agent. We believe that this type of reflection can help the Thinking Agent engage in effective reasoning. After removing the invalid intermediate reflection processes, we use \textless step\textgreater  to connect adjacent reasoning steps, thereby delineating each small step of reasoning. The reasoning process of the Thinking Agent falls between \textless thinking\textgreater and \textless /thinking\textgreater, while the reflection process of the Reflection Agent is enclosed within \textless  reflection\textgreater and \textless /reflection\textgreater. The entire code reasoning chain is encapsulated between \textless  ChainOfThought\textgreater and \textless /ChainOfThought\textgreater, with the final code answer provided thereafter. In the end, we synthesize the Code-COT dataset, which contains 2,810 high-quality code reasoning process data entries for SFT to get Cot-Coder model.

\textbf{SFT details} The base models we use are Qwen2.5-Coder-7B-Base and Qwen2.5-Coder-32B-Base. we use Code-COT dataset for supervised fine-tuning on the base models, obtaining the SFT models: COT-Coder-7B-SFT and COT-Coder-32B-SFT. We train the models for 3 epochs. The global batch size is set to 256, and the learning rate is set to 5e-6. The AdamW optimizer is used with a cosine decay learning rate scheduler. Weight\_decay is set to 0.1. $\beta_1$ is set to 0.9. $\beta_2$ is set to 0.95. Warmup\_steps is set to 30. DeepSpeed ZeRO2 is used to reduce GPU memory usage.

\textbf{self-explore\&self-improve} We employ the COT-Coder models to conduct tree search sampling, yielding a substantial number of reasoning steps, with each reasoning step treated as a node. Each node is scored based on the number of correct answers that can be derived from it. To ensure the exploration is efficient and effective, we have configured the sampling parameters for the tree search. Specifically, max\_path\_num is set to 5 and max\_depth\_num to 64. Additionally, we impose a maximum token limit of 25,000 for each individual path, truncating any paths that exceed this length. Any truncated path is considered an incorrect answer. Due to the existence of code problems that lack unit tests for correctness check, we employ Qwen2.5-7B-Instruct in combination with a compiler to generate test code and perform correctness check specifically for COT-Coder-7B-SFT. For COT-Coder-32B-SFT, we use Qwen2.5-32B-Instruct.
We then utilize the sampled data to conduct step-DPO. For each node, we take the reasoning process leading up to this node as the instruction and select the two child nodes with the greatest score difference to form a preference pair. 

For step-DPO, we train the models for 3 epochs. the learning\_rate is set to 5e-6. The global batch size is set to 256. The hyperparameter $\beta$ is set to 0.1. We use the AdamW optimizer and a cosine learning rate scheduler and the warmup\_ratio is set to 0.2. We add an additional NLL loss term with a scaling coefficient of 0.2 on the chosen sequence similar to the training approach used for Llama3. DeepSpeed ZeRO3 with CPU offload is used. 


For evaluation, we use the publicly available LiveCodeBench. LiveCodeBench is a challenging benchmark comprising 106 coding tasks, collected from July 1, 2024, to September 1, 2024, aimed at evaluating the code generation capabilities of LLMs. We follow the recommended setting by sampling 10 solutions for each problem with temperature as 0.2, and estimating the Pass@1 results.

The baselines we compared are the current powerful system-1 models. All results except those of our models are referenced from the official leaderboard.

\subsection{Main Results}
\begin{table}[h]
    \centering
    \caption{Main Results On LiveCodeBench}
    \resizebox{0.5\textwidth}{!}{
    \begin{tabular}{lcccc}
        \toprule
        Model  & Overall & Easy & Medium & Hard \\
        \midrule
        Claude-3.5-Sonnet-20241022 &35.4 &91.4  & 26.8  & 4.4  \\
        GPT-4O-2024-05-13 &33.6 &83.1  & 29.7  & 3.3  \\
        Gemini-Flash-2.0-Exp &30 &78.3  & 19.4  & 5.8 \\
        LLama3.3-70b-Instruct & 26.88 & 75.17 & 13.52 & 4.88 \\
        Deepseek-Coder-33B-Instruct & 19.62 & 57.24 & 10.88 & 1.16 \\
        Qwen2.5-Coder-32B-Instruct & 29.71 & 74.48 & 25.88 & 2.55 \\
        Deepseek-Coder-6.7B-Instruct & 13.58 & 37.58 & 10.29 & 0 \\
        Qwen2.5-Coder-7B-Instruct & 15.94 & 50.00 & 6.76 & 0.23 \\
        COT-Coder-7B-SFT(ours) & 20.18 & 54.48 & 15.29 & 0.93 \\
        COT-Coder-7B-StepDPO(ours) & 21.88 & 63.79 & 12.05 & 1.39 \\
        COT-Coder-32B-SFT(ours) & 33.49 & 81.72 & 29.11 & 4.41 \\
        COT-Coder-32B-StepDPO(ours) & 35.09 & 86.20 & 28.23 & 6.04 \\
        
        \bottomrule
    \end{tabular}
    \label{tab:five_column_table}
    }
\end{table}

Table \ref{tab:five_column_table} presents a comprehensive comparison of various models, encompassing both open-source and closed-source models. On LiveCodeBench, our CRPE-7B model scored 21, surpassing Qwen2.5-Coder-7B-Instruct and Deepseek-Coder-6.7B-Instruct. Our CRPE-32B model achieved a score of 34.22, which is comparable to mainstream models such as GPT-4o and Claude-3.5-Sonnet. The results show that our method effectively enhances the code generation capability of the LLM by improving its code reasoning ability.

\subsection{Analysis}
\subsubsection{Is cot data really useful?}
\begin{table}[ht]
    \centering
    \caption{COT VS Direct}
    \resizebox{0.5\textwidth}{!}{
    \begin{tabular}{lcccc}
        \toprule
        Model  & Overall & Easy & Medium & Hard \\
        \midrule
        COT-Coder-7B-SFT(partial data) & 19.15 & 55.51 & 11.76 & 0.46 \\
        Coder-7B-SFT-2 & 16.60 & 46.55 & 11.76 & 0.23 \\
        Coder-7B-SFT-1 & 14.71 & 45.17 & 7.35 & 0.0 \\ 
        
        \bottomrule
    \end{tabular}
    }
    \label{tab:five_column_table_1}
\end{table}

In this section, we aim to verify that incorporating reasoning process data into the SFT data can enhance the model's code generation capabilities. 

For the same coding problem, we use three different styles of answers as the data for SFT. The three different styles of answers are as follows: the correct answer written by a programmer, the correct answer generated by the LLM without the reasoning process, and the correct answer generated by the LLM with the reasoning process. Our coding problems and the answers with reasoning process are extracted from our Cot-Code-SFT Dataset. We then remove the intermediate reasoning process steps, retaining only the final answers to form the correct answers without reasoning process. Finally, we extract the correct answers written by programmers, which has been obtained during the data collection process. Some coding problems that do not contain correct answers written by programmers are removed to ensure that the coding problems used for training are the same. The difference between the code written by programmers and that generated by the LLM is that the answers written by programmers may lack comments and are more concise, while the code generated by the LLM includes line-by-line explanations, making it easier to understand and more aligned with the output style typical of LLMs. 

We use these three different styles of data to perform SFT based on the Qwen2.5-coder-7B-base model under the same training parameter settings, resulting in the following models: Coder-7B-SFT-1 (based on the correct answers written by programmers), Coder-7B-SFT-2 (based on the correct answers generated by the LLM without the reasoning process), and COT-Coder-7B-SFT(partial data) (based on the correct answers generated by the LLM with the reasoning process). 

\textbf{Training details}. We train the models for 3 epochs. The global batch size is set to 256, and the learning rate is set to 5e-6. We use the AdamW optimizer with a cosine decay learning rate scheduler, setting the warmup steps to 30. DeepSpeed ZeRO2 is used to reduce GPU memory usage. 

We then evaluate the three models on LiveCodeBench. Our experimental results indicate that the model trained on answers generated by LLMs (16.60) performes better than the model trained on correct answers written by programmers (14.71). This is also consistent with cases from some works on synthetic data, which indicate that using LLMs to rewrite collected answers leads to improved performance. The result of COT-Coder-7B-SFT and Coder-7B-SFT-2 shows that adding reasoning process steps further enhances the model's code generation capabilities(16.60 $\rightarrow$ 19.15). Therefore, in addition to rewriting answers with LLMs, synthesizing high-quality reasoning steps is also very valuable.

\subsubsection{Performance on HumanEval and MBPP}
In this section, we analyze the performance of our COT-Coder-7B-SFT on HumanEval and MBPP.

Humaneval and MBPP are also benchmarks used to evaluate the code generation capabilities of LLMs. However, unlike LiveCodeBench, which requires generating a complete solution for the code problem, HumanEval and MBPP primarily focus on function-level generation, specifically generating individual functions. Additionally, the names of the functions to be generated are already specified within the code problems. In LiveCodeBench, the descriptions of code problems are more detailed, clearly specifying the representation of inputs and outputs as well as the implementation requirements. In contrast, the descriptions of code problems in Humaneval and MBPP are relatively concise. The code problems in these two benchmarks are out-of-distribution for our COT-Coder. 
Our tests here use the code from EvalPlus\cite{liu2023codegeneratedchatgptreally}.

We use greedy sampling in our tests here. After the generation is complete, we extract the last code snippet generated for evaluation. Our COT-Coder-7B-SFT get 73.2/67.7 on HumanEval and 72.5/61.4 on MBPP.

In the results of HumanEval, COT-Coder-7B-SFT generates 44 incorrect answers. Among these, two answers fails due to incorrect function names. Specifically, the expected function names were special\_factorial and fibfib, but the generated names were brazilian\_factorial and fibfib\_iterative. If we fix the function names, COT-Coder's score on HumanEval improves to 74.4/68.9. The remaining issues primarily stem from errors in the generated code. The remaining failed cases are primarily due to issues in the code generated by COT-Coder. Some code problems in HumanEval with simple descriptions need to combine the example test cases for understanding, which are different from our SFT code instructions.

In the results of MBPP, COT-Coder-7B-SFT generates 104 incorrect answers, of which 32 answers could not correctly extract the function code. In these 32 answers, 30 answers have incorrect function names, and 2 answers are in a loop of thinking and reflection, resulting in the generated text exceeding the max\_tokens limit. We analyze the 30 incorrect function names and find that the original function names don't adhere to naming conventions, which leads the model to modify the function names when generating answers. For example, the original function name is check\_Consecutive, and the model modifies it to check\_consecutive. Although the modified function names comply better with naming conventions and the code remains correct, this change ultimately leads to test failures.
After correcting the 30 incorrect function names, COT-Coder-7B-SFT's score on MBPP improves to 78.6/66.1. 

\section{conclusion}
In this paper, we introduce CRPE, a method that enhances the code generation capabilities of LLMs by explicitly augmenting their code reasoning abilities. In brief, it is a three-stage data synthesis
and training method, leveraging existing powerful system-1 models to enhance the code reasoning capabilities of code reasoning models, finally achieving self-improvement of the code reasoning models. We also produce a high-quality code problem dataset, and a high-quality code reasoning process dataset. Experimental results on the LiveCodeBench demonstrate the effectiveness of CRPE in enhancing the code generation capabilities of LLMs. Future work will focus on improving the efficiency of self-improvement of the base models and optimizing the token length during model inference.

\section{limitations}
CRPE enhances the code generation capabilities of LLMs by reinforcing their code reasoning abilities, but it also faces new challenges. 

Firstly, this method relies on high-quality code problems. Although there is currently a considerable amount of work related to synthetic data, these efforts do not construct data tailored to the model being trained. In addition, there is a lack of methods to evaluate the effectiveness of the data for the model being trained, which may result in a waste of resources. 

Secondly, our CRPE models may get trapped in a feedback loop during inference, failing to provide a final answer. This is mainly because the reasoning process consists of reasoning part and reflection part, and the model may repeatedly cycle through reasoning and reflection, unable to arrive at a final answer. Future iterations will address this issue. By addressing these limitations, we can make our CRPE method more efficient and practical.
\bibliographystyle{IEEEtran}
\bibliography{ref}

\begin{thebibliography}{10}
\providecommand{\url}[1]{#1}
\csname url@samestyle\endcsname
\providecommand{\newblock}{\relax}
\providecommand{\bibinfo}[2]{#2}
\providecommand{\BIBentrySTDinterwordspacing}{\spaceskip=0pt\relax}
\providecommand{\BIBentryALTinterwordstretchfactor}{4}
\providecommand{\BIBentryALTinterwordspacing}{\spaceskip=\fontdimen2\font plus
\BIBentryALTinterwordstretchfactor\fontdimen3\font minus \fontdimen4\font\relax}
\providecommand{\BIBforeignlanguage}[2]{{%
\expandafter\ifx\csname l@#1\endcsname\relax
\typeout{** WARNING: IEEEtran.bst: No hyphenation pattern has been}%
\typeout{** loaded for the language `#1'. Using the pattern for}%
\typeout{** the default language instead.}%
\else
\language=\csname l@#1\endcsname
\fi
#2}}
\providecommand{\BIBdecl}{\relax}
\BIBdecl

\bibitem{hui_qwen25-coder_2024}
\BIBentryALTinterwordspacing
B.~Hui, J.~Yang, Z.~Cui, J.~Yang, D.~Liu, L.~Zhang, T.~Liu, J.~Zhang, B.~Yu, K.~Dang, A.~Yang, R.~Men, F.~Huang, X.~Ren, X.~Ren, J.~Zhou, and J.~Lin, ``\BIBforeignlanguage{en}{Qwen2.5-{Coder} {Technical} {Report}},'' Sep. 2024, arXiv:2409.12186 [cs]. [Online]. Available: \url{http://arxiv.org/abs/2409.12186}
\BIBentrySTDinterwordspacing

\bibitem{deepseek-ai_deepseek-coder-v2_2024}
\BIBentryALTinterwordspacing
DeepSeek-AI, Q.~Zhu, D.~Guo, Z.~Shao, D.~Yang, P.~Wang, R.~Xu, Y.~Wu, Y.~Li, H.~Gao, S.~Ma, W.~Zeng, X.~Bi, Z.~Gu, H.~Xu, D.~Dai, K.~Dong, L.~Zhang, Y.~Piao, Z.~Gou, Z.~Xie, Z.~Hao, B.~Wang, J.~Song, D.~Chen, X.~Xie, K.~Guan, Y.~You, A.~Liu, Q.~Du, W.~Gao, X.~Lu, Q.~Chen, Y.~Wang, C.~Deng, J.~Li, C.~Zhao, C.~Ruan, F.~Luo, and W.~Liang, ``\BIBforeignlanguage{en}{{DeepSeek}-{Coder}-{V2}: {Breaking} the {Barrier} of {Closed}-{Source} {Models} in {Code} {Intelligence}},'' Jun. 2024, arXiv:2406.11931 [cs]. [Online]. Available: \url{http://arxiv.org/abs/2406.11931}
\BIBentrySTDinterwordspacing

\bibitem{Wei-CoT-2201-11903}
\BIBentryALTinterwordspacing
J.~Wei, X.~Wang, D.~Schuurmans, M.~Bosma, E.~H. Chi, Q.~Le, and D.~Zhou, ``Chain of thought prompting elicits reasoning in large language models,'' \emph{CoRR}, vol. abs/2201.11903, 2022. [Online]. Available: \url{https://arxiv.org/abs/2201.11903}
\BIBentrySTDinterwordspacing

\bibitem{yao2023treethoughtsdeliberateproblem}
\BIBentryALTinterwordspacing
S.~Yao, D.~Yu, J.~Zhao, I.~Shafran, T.~L. Griffiths, Y.~Cao, and K.~Narasimhan, ``Tree of thoughts: Deliberate problem solving with large language models,'' 2023. [Online]. Available: \url{https://arxiv.org/abs/2305.10601}
\BIBentrySTDinterwordspacing

\bibitem{chen2024alphamathzeroprocesssupervision}
\BIBentryALTinterwordspacing
G.~Chen, M.~Liao, C.~Li, and K.~Fan, ``Alphamath almost zero: Process supervision without process,'' 2024. [Online]. Available: \url{https://arxiv.org/abs/2405.03553}
\BIBentrySTDinterwordspacing

\bibitem{zhang2024llamaberrypairwiseoptimizationo1like}
\BIBentryALTinterwordspacing
D.~Zhang, J.~Wu, J.~Lei, T.~Che, J.~Li, T.~Xie, X.~Huang, S.~Zhang, M.~Pavone, Y.~Li, W.~Ouyang, and D.~Zhou, ``Llama-berry: Pairwise optimization for o1-like olympiad-level mathematical reasoning,'' 2024. [Online]. Available: \url{https://arxiv.org/abs/2410.02884}
\BIBentrySTDinterwordspacing

\bibitem{luo2023wizardcoderempoweringcodelarge}
\BIBentryALTinterwordspacing
Z.~Luo, C.~Xu, P.~Zhao, Q.~Sun, X.~Geng, W.~Hu, C.~Tao, J.~Ma, Q.~Lin, and D.~Jiang, ``Wizardcoder: Empowering code large language models with evol-instruct,'' 2023. [Online]. Available: \url{https://arxiv.org/abs/2306.08568}
\BIBentrySTDinterwordspacing

\bibitem{yu2024wavecoderwidespreadversatileenhancement}
\BIBentryALTinterwordspacing
Z.~Yu, X.~Zhang, N.~Shang, Y.~Huang, C.~Xu, Y.~Zhao, W.~Hu, and Q.~Yin, ``Wavecoder: Widespread and versatile enhancement for code large language models by instruction tuning,'' 2024. [Online]. Available: \url{https://arxiv.org/abs/2312.14187}
\BIBentrySTDinterwordspacing

\bibitem{wei2024magicoderempoweringcodegeneration}
\BIBentryALTinterwordspacing
Y.~Wei, Z.~Wang, J.~Liu, Y.~Ding, and L.~Zhang, ``Magicoder: Empowering code generation with oss-instruct,'' 2024. [Online]. Available: \url{https://arxiv.org/abs/2312.02120}
\BIBentrySTDinterwordspacing

\bibitem{dou2024stepcoderimprovecodegeneration}
\BIBentryALTinterwordspacing
S.~Dou, Y.~Liu, H.~Jia, L.~Xiong, E.~Zhou, W.~Shen, J.~Shan, C.~Huang, X.~Wang, X.~Fan, Z.~Xi, Y.~Zhou, T.~Ji, R.~Zheng, Q.~Zhang, X.~Huang, and T.~Gui, ``Stepcoder: Improve code generation with reinforcement learning from compiler feedback,'' 2024. [Online]. Available: \url{https://arxiv.org/abs/2402.01391}
\BIBentrySTDinterwordspacing

\bibitem{shojaee2023executionbasedcodegenerationusing}
\BIBentryALTinterwordspacing
P.~Shojaee, A.~Jain, S.~Tipirneni, and C.~K. Reddy, ``Execution-based code generation using deep reinforcement learning,'' 2023. [Online]. Available: \url{https://arxiv.org/abs/2301.13816}
\BIBentrySTDinterwordspacing

\bibitem{liu2023rltfreinforcementlearningunit}
\BIBentryALTinterwordspacing
J.~Liu, Y.~Zhu, K.~Xiao, Q.~Fu, X.~Han, W.~Yang, and D.~Ye, ``Rltf: Reinforcement learning from unit test feedback,'' 2023. [Online]. Available: \url{https://arxiv.org/abs/2307.04349}
\BIBentrySTDinterwordspacing

\bibitem{hui2024qwen25codertechnicalreport}
\BIBentryALTinterwordspacing
B.~Hui, J.~Yang, Z.~Cui, J.~Yang, D.~Liu, L.~Zhang, T.~Liu, J.~Zhang, B.~Yu, K.~Lu, K.~Dang, Y.~Fan, Y.~Zhang, A.~Yang, R.~Men, F.~Huang, B.~Zheng, Y.~Miao, S.~Quan, Y.~Feng, X.~Ren, X.~Ren, J.~Zhou, and J.~Lin, ``Qwen2.5-coder technical report,'' 2024. [Online]. Available: \url{https://arxiv.org/abs/2409.12186}
\BIBentrySTDinterwordspacing

\bibitem{guo2024deepseekcoderlargelanguagemodel}
\BIBentryALTinterwordspacing
D.~Guo, Q.~Zhu, D.~Yang, Z.~Xie, K.~Dong, W.~Zhang, G.~Chen, X.~Bi, Y.~Wu, Y.~K. Li, F.~Luo, Y.~Xiong, and W.~Liang, ``Deepseek-coder: When the large language model meets programming -- the rise of code intelligence,'' 2024. [Online]. Available: \url{https://arxiv.org/abs/2401.14196}
\BIBentrySTDinterwordspacing

\bibitem{rozière2024codellamaopenfoundation}
\BIBentryALTinterwordspacing
B.~Rozière, J.~Gehring, F.~Gloeckle, S.~Sootla, I.~Gat, X.~E. Tan, Y.~Adi, J.~Liu, R.~Sauvestre, T.~Remez, J.~Rapin, A.~Kozhevnikov, I.~Evtimov, J.~Bitton, M.~Bhatt, C.~C. Ferrer, A.~Grattafiori, W.~Xiong, A.~Défossez, J.~Copet, F.~Azhar, H.~Touvron, L.~Martin, N.~Usunier, T.~Scialom, and G.~Synnaeve, ``Code llama: Open foundation models for code,'' 2024. [Online]. Available: \url{https://arxiv.org/abs/2308.12950}
\BIBentrySTDinterwordspacing

\bibitem{grattafiori2024llama3herdmodels}
\BIBentryALTinterwordspacing
A.~Grattafiori, A.~Dubey, A.~Jauhri, A.~Pandey, A.~Kadian, A.~Al-Dahle, A.~Letman, A.~Mathur, A.~Schelten, A.~Vaughan, A.~Yang, A.~Fan, A.~Goyal, A.~Hartshorn, A.~Yang, A.~Mitra, A.~Sravankumar, A.~Korenev, A.~Hinsvark, A.~Rao, A.~Zhang, A.~Rodriguez, A.~Gregerson, A.~Spataru, B.~Roziere, B.~Biron, B.~Tang, B.~Chern, C.~Caucheteux, C.~Nayak, C.~Bi, C.~Marra, C.~McConnell, C.~Keller, C.~Touret, C.~Wu, C.~Wong, C.~C. Ferrer, C.~Nikolaidis, D.~Allonsius, D.~Song, D.~Pintz, D.~Livshits, D.~Wyatt, D.~Esiobu, D.~Choudhary, D.~Mahajan, D.~Garcia-Olano, D.~Perino, D.~Hupkes, E.~Lakomkin, E.~AlBadawy, E.~Lobanova, E.~Dinan, E.~M. Smith, F.~Radenovic, F.~Guzmán, F.~Zhang, G.~Synnaeve, G.~Lee, G.~L. Anderson, G.~Thattai, G.~Nail, G.~Mialon, G.~Pang, G.~Cucurell, H.~Nguyen, H.~Korevaar, H.~Xu, H.~Touvron, I.~Zarov, I.~A. Ibarra, I.~Kloumann, I.~Misra, I.~Evtimov, J.~Zhang, J.~Copet, J.~Lee, J.~Geffert, J.~Vranes, J.~Park, J.~Mahadeokar, J.~Shah, J.~van~der Linde, J.~Billock, J.~Hong, J.~Lee, J.~Fu, J.~Chi, J.~Huang,
  J.~Liu, J.~Wang, J.~Yu, J.~Bitton, J.~Spisak, J.~Park, J.~Rocca, J.~Johnstun, J.~Saxe, J.~Jia, K.~V. Alwala, K.~Prasad, K.~Upasani, K.~Plawiak, K.~Li, K.~Heafield, K.~Stone, K.~El-Arini, K.~Iyer, K.~Malik, K.~Chiu, K.~Bhalla, K.~Lakhotia, L.~Rantala-Yeary, L.~van~der Maaten, L.~Chen, L.~Tan, L.~Jenkins, L.~Martin, L.~Madaan, L.~Malo, L.~Blecher, L.~Landzaat, L.~de~Oliveira, M.~Muzzi, M.~Pasupuleti, M.~Singh, M.~Paluri, M.~Kardas, M.~Tsimpoukelli, M.~Oldham, M.~Rita, M.~Pavlova, M.~Kambadur, M.~Lewis, M.~Si, M.~K. Singh, M.~Hassan, N.~Goyal, N.~Torabi, N.~Bashlykov, N.~Bogoychev, N.~Chatterji, N.~Zhang, O.~Duchenne, O.~Çelebi, P.~Alrassy, P.~Zhang, P.~Li, P.~Vasic, P.~Weng, P.~Bhargava, P.~Dubal, P.~Krishnan, P.~S. Koura, P.~Xu, Q.~He, Q.~Dong, R.~Srinivasan, R.~Ganapathy, R.~Calderer, R.~S. Cabral, R.~Stojnic, R.~Raileanu, R.~Maheswari, R.~Girdhar, R.~Patel, R.~Sauvestre, R.~Polidoro, R.~Sumbaly, R.~Taylor, R.~Silva, R.~Hou, R.~Wang, S.~Hosseini, S.~Chennabasappa, S.~Singh, S.~Bell, S.~S. Kim, S.~Edunov,
  S.~Nie, S.~Narang, S.~Raparthy, S.~Shen, S.~Wan, S.~Bhosale, S.~Zhang, S.~Vandenhende, S.~Batra, S.~Whitman, S.~Sootla, S.~Collot, S.~Gururangan, S.~Borodinsky, T.~Herman, T.~Fowler, T.~Sheasha, T.~Georgiou, T.~Scialom, T.~Speckbacher, T.~Mihaylov, T.~Xiao, U.~Karn, V.~Goswami, V.~Gupta, V.~Ramanathan, V.~Kerkez, V.~Gonguet, V.~Do, V.~Vogeti, V.~Albiero, V.~Petrovic, W.~Chu, W.~Xiong, W.~Fu, W.~Meers, X.~Martinet, X.~Wang, X.~Wang, X.~E. Tan, X.~Xia, X.~Xie, X.~Jia, X.~Wang, Y.~Goldschlag, Y.~Gaur, Y.~Babaei, Y.~Wen, Y.~Song, Y.~Zhang, Y.~Li, Y.~Mao, Z.~D. Coudert, Z.~Yan, Z.~Chen, Z.~Papakipos, A.~Singh, A.~Srivastava, A.~Jain, A.~Kelsey, A.~Shajnfeld, A.~Gangidi, A.~Victoria, A.~Goldstand, A.~Menon, A.~Sharma, A.~Boesenberg, A.~Baevski, A.~Feinstein, A.~Kallet, A.~Sangani, A.~Teo, A.~Yunus, A.~Lupu, A.~Alvarado, A.~Caples, A.~Gu, A.~Ho, A.~Poulton, A.~Ryan, A.~Ramchandani, A.~Dong, A.~Franco, A.~Goyal, A.~Saraf, A.~Chowdhury, A.~Gabriel, A.~Bharambe, A.~Eisenman, A.~Yazdan, B.~James, B.~Maurer,
  B.~Leonhardi, B.~Huang, B.~Loyd, B.~D. Paola, B.~Paranjape, B.~Liu, B.~Wu, B.~Ni, B.~Hancock, B.~Wasti, B.~Spence, B.~Stojkovic, B.~Gamido, B.~Montalvo, C.~Parker, C.~Burton, C.~Mejia, C.~Liu, C.~Wang, C.~Kim, C.~Zhou, C.~Hu, C.-H. Chu, C.~Cai, C.~Tindal, C.~Feichtenhofer, C.~Gao, D.~Civin, D.~Beaty, D.~Kreymer, D.~Li, D.~Adkins, D.~Xu, D.~Testuggine, D.~David, D.~Parikh, D.~Liskovich, D.~Foss, D.~Wang, D.~Le, D.~Holland, E.~Dowling, E.~Jamil, E.~Montgomery, E.~Presani, E.~Hahn, E.~Wood, E.-T. Le, E.~Brinkman, E.~Arcaute, E.~Dunbar, E.~Smothers, F.~Sun, F.~Kreuk, F.~Tian, F.~Kokkinos, F.~Ozgenel, F.~Caggioni, F.~Kanayet, F.~Seide, G.~M. Florez, G.~Schwarz, G.~Badeer, G.~Swee, G.~Halpern, G.~Herman, G.~Sizov, Guangyi, Zhang, G.~Lakshminarayanan, H.~Inan, H.~Shojanazeri, H.~Zou, H.~Wang, H.~Zha, H.~Habeeb, H.~Rudolph, H.~Suk, H.~Aspegren, H.~Goldman, H.~Zhan, I.~Damlaj, I.~Molybog, I.~Tufanov, I.~Leontiadis, I.-E. Veliche, I.~Gat, J.~Weissman, J.~Geboski, J.~Kohli, J.~Lam, J.~Asher, J.-B. Gaya, J.~Marcus,
  J.~Tang, J.~Chan, J.~Zhen, J.~Reizenstein, J.~Teboul, J.~Zhong, J.~Jin, J.~Yang, J.~Cummings, J.~Carvill, J.~Shepard, J.~McPhie, J.~Torres, J.~Ginsburg, J.~Wang, K.~Wu, K.~H. U, K.~Saxena, K.~Khandelwal, K.~Zand, K.~Matosich, K.~Veeraraghavan, K.~Michelena, K.~Li, K.~Jagadeesh, K.~Huang, K.~Chawla, K.~Huang, L.~Chen, L.~Garg, L.~A, L.~Silva, L.~Bell, L.~Zhang, L.~Guo, L.~Yu, L.~Moshkovich, L.~Wehrstedt, M.~Khabsa, M.~Avalani, M.~Bhatt, M.~Mankus, M.~Hasson, M.~Lennie, M.~Reso, M.~Groshev, M.~Naumov, M.~Lathi, M.~Keneally, M.~Liu, M.~L. Seltzer, M.~Valko, M.~Restrepo, M.~Patel, M.~Vyatskov, M.~Samvelyan, M.~Clark, M.~Macey, M.~Wang, M.~J. Hermoso, M.~Metanat, M.~Rastegari, M.~Bansal, N.~Santhanam, N.~Parks, N.~White, N.~Bawa, N.~Singhal, N.~Egebo, N.~Usunier, N.~Mehta, N.~P. Laptev, N.~Dong, N.~Cheng, O.~Chernoguz, O.~Hart, O.~Salpekar, O.~Kalinli, P.~Kent, P.~Parekh, P.~Saab, P.~Balaji, P.~Rittner, P.~Bontrager, P.~Roux, P.~Dollar, P.~Zvyagina, P.~Ratanchandani, P.~Yuvraj, Q.~Liang, R.~Alao, R.~Rodriguez,
  R.~Ayub, R.~Murthy, R.~Nayani, R.~Mitra, R.~Parthasarathy, R.~Li, R.~Hogan, R.~Battey, R.~Wang, R.~Howes, R.~Rinott, S.~Mehta, S.~Siby, S.~J. Bondu, S.~Datta, S.~Chugh, S.~Hunt, S.~Dhillon, S.~Sidorov, S.~Pan, S.~Mahajan, S.~Verma, S.~Yamamoto, S.~Ramaswamy, S.~Lindsay, S.~Lindsay, S.~Feng, S.~Lin, S.~C. Zha, S.~Patil, S.~Shankar, S.~Zhang, S.~Zhang, S.~Wang, S.~Agarwal, S.~Sajuyigbe, S.~Chintala, S.~Max, S.~Chen, S.~Kehoe, S.~Satterfield, S.~Govindaprasad, S.~Gupta, S.~Deng, S.~Cho, S.~Virk, S.~Subramanian, S.~Choudhury, S.~Goldman, T.~Remez, T.~Glaser, T.~Best, T.~Koehler, T.~Robinson, T.~Li, T.~Zhang, T.~Matthews, T.~Chou, T.~Shaked, V.~Vontimitta, V.~Ajayi, V.~Montanez, V.~Mohan, V.~S. Kumar, V.~Mangla, V.~Ionescu, V.~Poenaru, V.~T. Mihailescu, V.~Ivanov, W.~Li, W.~Wang, W.~Jiang, W.~Bouaziz, W.~Constable, X.~Tang, X.~Wu, X.~Wang, X.~Wu, X.~Gao, Y.~Kleinman, Y.~Chen, Y.~Hu, Y.~Jia, Y.~Qi, Y.~Li, Y.~Zhang, Y.~Zhang, Y.~Adi, Y.~Nam, Yu, Wang, Y.~Zhao, Y.~Hao, Y.~Qian, Y.~Li, Y.~He, Z.~Rait, Z.~DeVito,
  Z.~Rosnbrick, Z.~Wen, Z.~Yang, Z.~Zhao, and Z.~Ma, ``The llama 3 herd of models,'' 2024. [Online]. Available: \url{https://arxiv.org/abs/2407.21783}
\BIBentrySTDinterwordspacing

\bibitem{lozhkov2024starcoder2stackv2}
\BIBentryALTinterwordspacing
A.~Lozhkov, R.~Li, L.~B. Allal, F.~Cassano, J.~Lamy-Poirier, N.~Tazi, A.~Tang, D.~Pykhtar, J.~Liu, Y.~Wei, T.~Liu, M.~Tian, D.~Kocetkov, A.~Zucker, Y.~Belkada, Z.~Wang, Q.~Liu, D.~Abulkhanov, I.~Paul, Z.~Li, W.-D. Li, M.~Risdal, J.~Li, J.~Zhu, T.~Y. Zhuo, E.~Zheltonozhskii, N.~O.~O. Dade, W.~Yu, L.~Krauß, N.~Jain, Y.~Su, X.~He, M.~Dey, E.~Abati, Y.~Chai, N.~Muennighoff, X.~Tang, M.~Oblokulov, C.~Akiki, M.~Marone, C.~Mou, M.~Mishra, A.~Gu, B.~Hui, T.~Dao, A.~Zebaze, O.~Dehaene, N.~Patry, C.~Xu, J.~McAuley, H.~Hu, T.~Scholak, S.~Paquet, J.~Robinson, C.~J. Anderson, N.~Chapados, M.~Patwary, N.~Tajbakhsh, Y.~Jernite, C.~M. Ferrandis, L.~Zhang, S.~Hughes, T.~Wolf, A.~Guha, L.~von Werra, and H.~de~Vries, ``Starcoder 2 and the stack v2: The next generation,'' 2024. [Online]. Available: \url{https://arxiv.org/abs/2402.19173}
\BIBentrySTDinterwordspacing

\bibitem{wang2022compilableneuralcodegeneration}
\BIBentryALTinterwordspacing
X.~Wang, Y.~Wang, Y.~Wan, F.~Mi, Y.~Li, P.~Zhou, J.~Liu, H.~Wu, X.~Jiang, and Q.~Liu, ``Compilable neural code generation with compiler feedback,'' 2022. [Online]. Available: \url{https://arxiv.org/abs/2203.05132}
\BIBentrySTDinterwordspacing

\bibitem{le2022coderlmasteringcodegeneration}
\BIBentryALTinterwordspacing
H.~Le, Y.~Wang, A.~D. Gotmare, S.~Savarese, and S.~C.~H. Hoi, ``Coderl: Mastering code generation through pretrained models and deep reinforcement learning,'' 2022. [Online]. Available: \url{https://arxiv.org/abs/2207.01780}
\BIBentrySTDinterwordspacing

\bibitem{zhang2024codedpoaligningcodemodels}
\BIBentryALTinterwordspacing
K.~Zhang, G.~Li, Y.~Dong, J.~Xu, J.~Zhang, J.~Su, Y.~Liu, and Z.~Jin, ``Codedpo: Aligning code models with self generated and verified source code,'' 2024. [Online]. Available: \url{https://arxiv.org/abs/2410.05605}
\BIBentrySTDinterwordspacing

\bibitem{openaio1}
\BIBentryALTinterwordspacing
OpenAI, ``Learning to reason with large language models.'' 2024. [Online]. Available: \url{https://openai.com/index/learning-to-reason-with-llms/}
\BIBentrySTDinterwordspacing

\bibitem{qwenqwq}
\BIBentryALTinterwordspacing
Q.~Team, ``Qwq-32b-preview.'' 2024. [Online]. Available: \url{https://qwenlm.github.io/zh/blog/qwq-32b-preview/}
\BIBentrySTDinterwordspacing

\bibitem{dai2024processsupervisionguidedpolicyoptimization}
\BIBentryALTinterwordspacing
N.~Dai, Z.~Wu, R.~Zheng, Z.~Wei, W.~Shi, X.~Jin, G.~Liu, C.~Dun, L.~Huang, and L.~Yan, ``Process supervision-guided policy optimization for code generation,'' 2024. [Online]. Available: \url{https://arxiv.org/abs/2410.17621}
\BIBentrySTDinterwordspacing

\bibitem{zhang2024o1codero1replicationcoding}
\BIBentryALTinterwordspacing
Y.~Zhang, S.~Wu, Y.~Yang, J.~Shu, J.~Xiao, C.~Kong, and J.~Sang, ``o1-coder: an o1 replication for coding,'' 2024. [Online]. Available: \url{https://arxiv.org/abs/2412.00154}
\BIBentrySTDinterwordspacing

\bibitem{rafailov2024directpreferenceoptimizationlanguage}
\BIBentryALTinterwordspacing
R.~Rafailov, A.~Sharma, E.~Mitchell, S.~Ermon, C.~D. Manning, and C.~Finn, ``Direct preference optimization: Your language model is secretly a reward model,'' 2024. [Online]. Available: \url{https://arxiv.org/abs/2305.18290}
\BIBentrySTDinterwordspacing

\bibitem{lai2024stepdpostepwisepreferenceoptimization}
\BIBentryALTinterwordspacing
X.~Lai, Z.~Tian, Y.~Chen, S.~Yang, X.~Peng, and J.~Jia, ``Step-dpo: Step-wise preference optimization for long-chain reasoning of llms,'' 2024. [Online]. Available: \url{https://arxiv.org/abs/2406.18629}
\BIBentrySTDinterwordspacing

\bibitem{xin2024deepseekproverv15harnessingproofassistant}
\BIBentryALTinterwordspacing
H.~Xin, Z.~Z. Ren, J.~Song, Z.~Shao, W.~Zhao, H.~Wang, B.~Liu, L.~Zhang, X.~Lu, Q.~Du, W.~Gao, Q.~Zhu, D.~Yang, Z.~Gou, Z.~F. Wu, F.~Luo, and C.~Ruan, ``Deepseek-prover-v1.5: Harnessing proof assistant feedback for reinforcement learning and monte-carlo tree search,'' 2024. [Online]. Available: \url{https://arxiv.org/abs/2408.08152}
\BIBentrySTDinterwordspacing

\bibitem{liu2023codegeneratedchatgptreally}
\BIBentryALTinterwordspacing
J.~Liu, C.~S. Xia, Y.~Wang, and L.~Zhang, ``Is your code generated by chatgpt really correct? rigorous evaluation of large language models for code generation,'' 2023. [Online]. Available: \url{https://arxiv.org/abs/2305.01210}
\BIBentrySTDinterwordspacing

\end{thebibliography}

\newpage
\appendix
\onecolumn
\section{Prompt Details.}
\begin{figure}[h]
    \centering
    \includegraphics[width=0.8\textwidth]{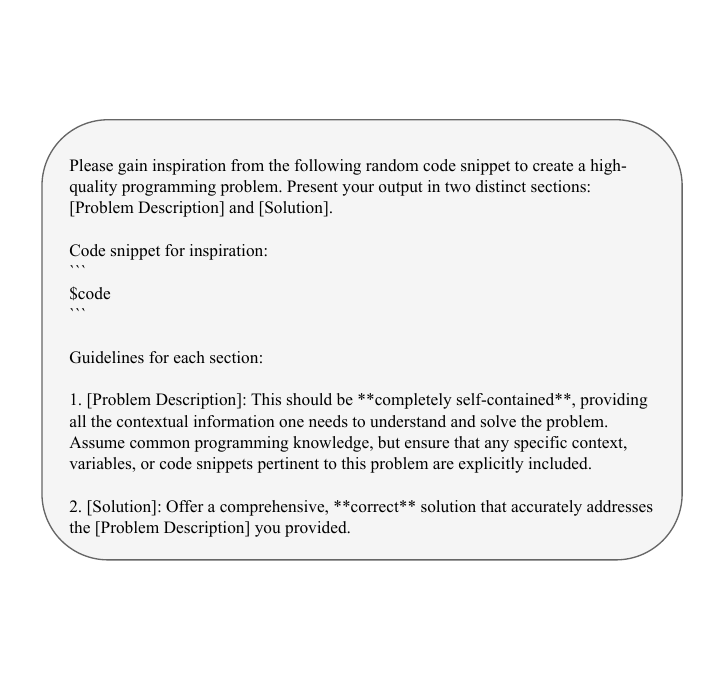}
    \caption{Prompt for generating code problem} 
    \label{fig:my_label_problem_generate} 
\end{figure}

\begin{figure}[h]
    \centering
    \includegraphics[width=0.8\textwidth]{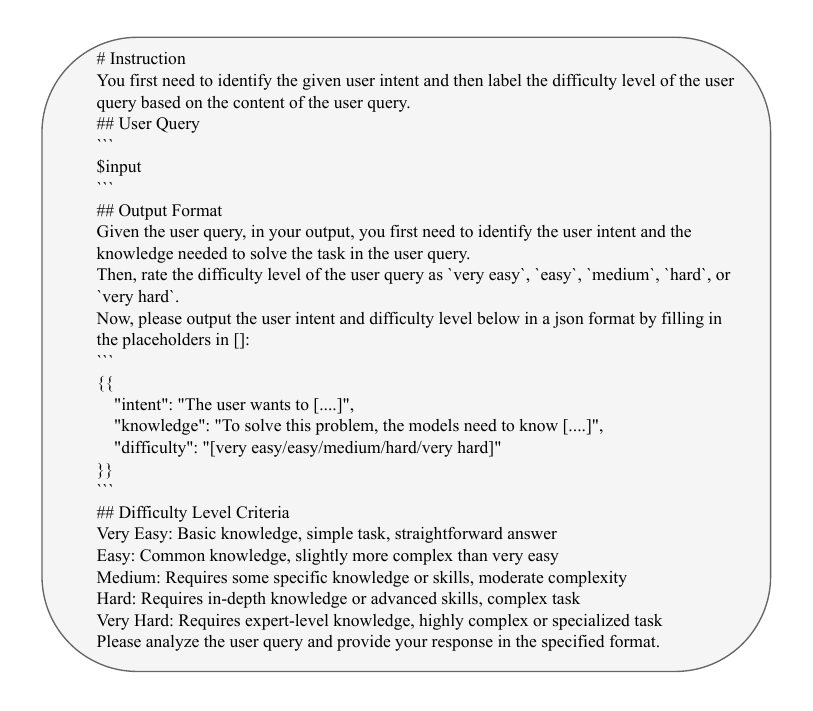}
    \caption{Prompt for difficulty analysis} 
    \label{fig:prompt_to_diff_analysis} 
\end{figure}

\begin{figure}[h]
    \centering
    \includegraphics[width=0.8\textwidth]{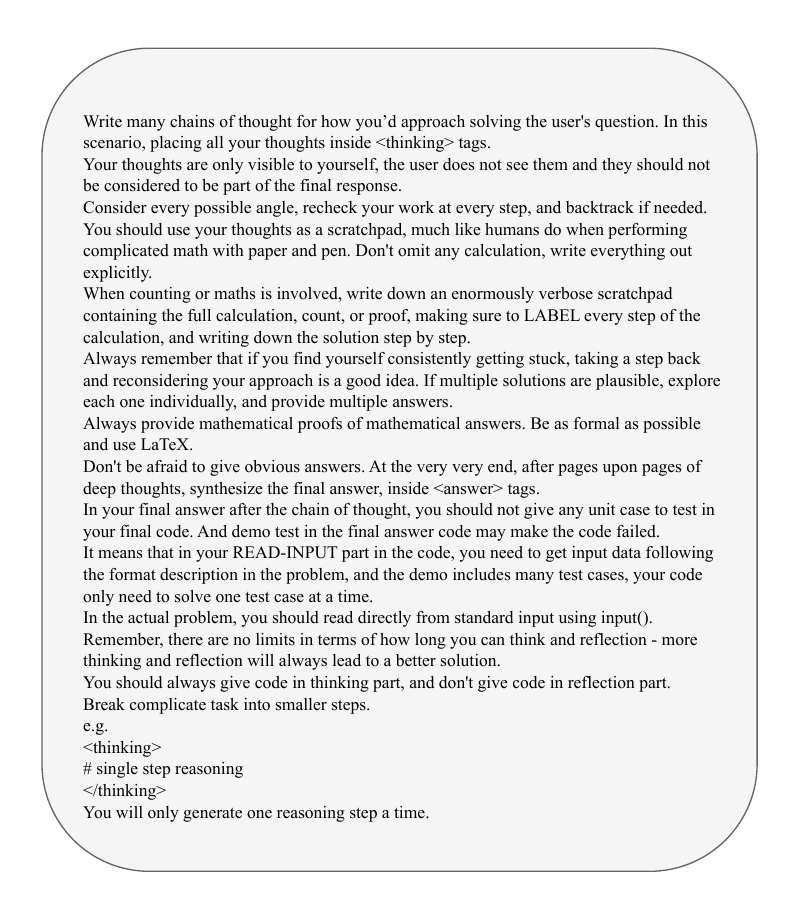}
    \caption{Prompt for Thinking Agent} 
    \label{fig:prompt_to_thinking_agent} 
\end{figure}

\begin{figure}[h]
    \centering
    \includegraphics[width=0.8\textwidth]{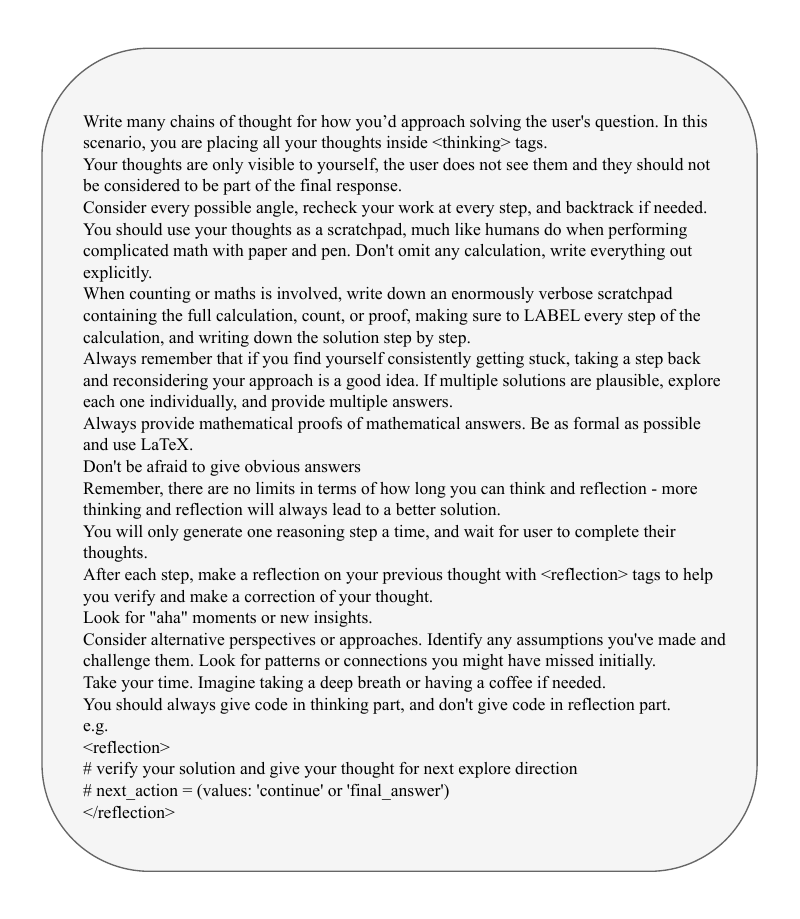}
    \caption{Prompt for Reflection Agent} 
    \label{fig:prompt_to_reflection_agent} 
\end{figure}

\begin{figure}[h]
    \centering
    \begin{small}
        \includegraphics[width=0.8\textwidth, height=0.8\textheight]{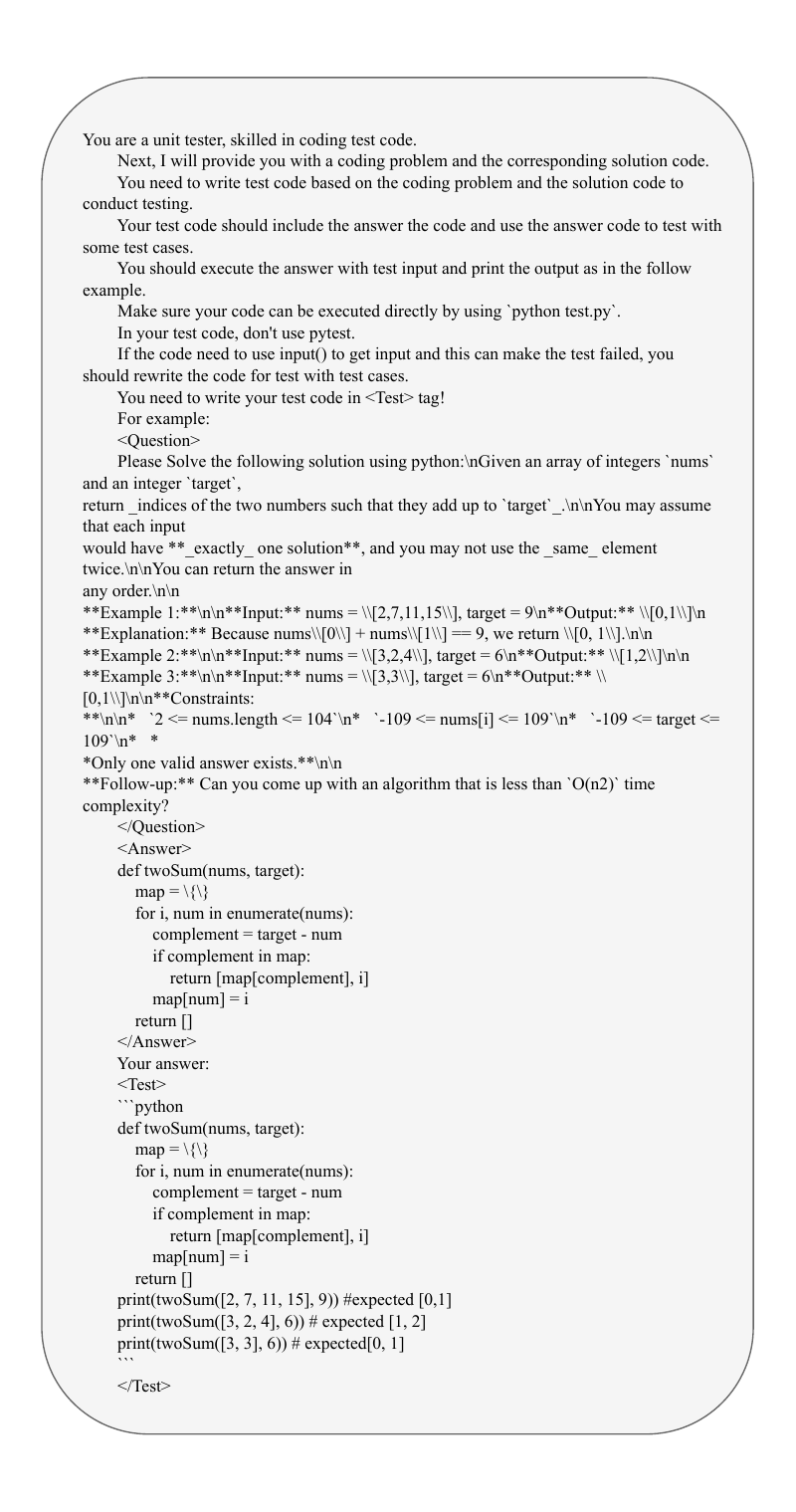}
    \caption{Prompt for Execution Agent to generate test code} 
    \label{fig:prompt_to_execution_agent_generate_test} 
    \end{small}
\end{figure}

\begin{figure}[h]
    \centering
    \includegraphics[width=0.8\textwidth]{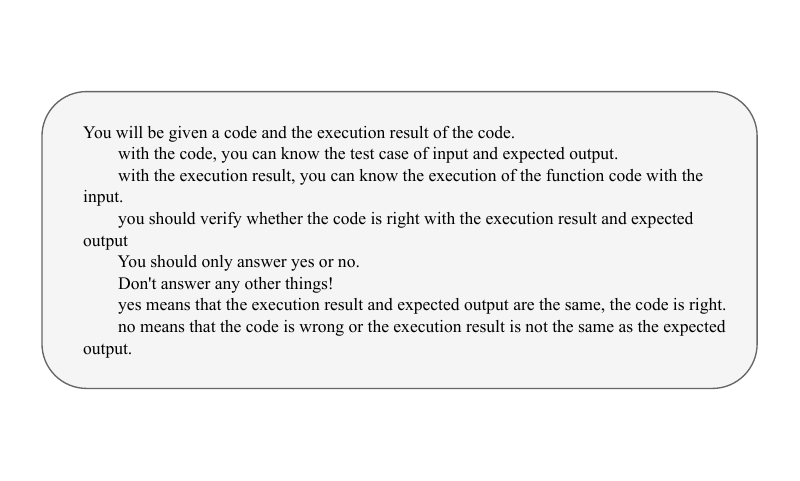}
    \caption{Prompt for Execution Agent to check correctness of execution result} 
    \label{fig:prompt_to_execution_agent_check_correct} 
\end{figure}

\end{document}